# Orbital-angular-momentum dependent speckles for spatial mode sorting and multiplexed data transmission

RUI MA,[1,2] KE HAI LUO,[1] ZHAO WANG,[3] JING SONG HE,[2] WEI LI ZHANG,[3] DIAN YUAN FAN,[1] ANDERSON S. L. GOMES,[4] AND JUN LIU[1,*]

[1]*International Collaborative Laboratory of 2D Materials for Optoelectronics Science and Technology of Ministry of Education, Institute of Microscale Optoelectronics, Shenzhen University, Shenzhen 518060, China*
[2]*Institute for Advanced Study, Shenzhen University, Shenzhen 518060, China*
[3]*School of Information and Communication Engineering, University of Electronic Science & Technology of China, Chengdu 611731, China*
[4]*Departamento de Física, Universidade Federal de Pernambuco, 50670-901 Recife-PE, Brazil*
*\*liuj1987@szu.edu.cn*

**Abstract:** Characterizing the orbital angular momentum (OAM) of a vortex beam is critically important for OAM-encoded data transfer. However, in typical OAM-based applications where vortex beams transmit through diffusers, the accompanying scattering effect tends to be either deliberately prevented, or characterized and then modulated actively based on complex wavefront shaping and interferometry techniques. Here, we aim to investigate the characteristics of blurred speckles obtained after a vortex beam transmits through a ground glass diffuser. It is theoretically and experimentally demonstrated that a cross-correlation annulus can be identified by implementing the cross-correlation operation between speckle patterns corresponding to vortex beams with different OAM values. Besides, it is worth noting that, the size of the cross-correlation annulus is determined by the absolute value of the topological charge difference between the two corresponding vortex beams. Based on this mechanism, the OAM modes can be easily sorted from the incoherently measured OAM-dependent speckles as well as their cross-correlation. Furthermore, to make full use of the orthogonal feature of the OAM-dependent speckles, demultiplexing of OAM-encoded data transfer is verified using a ground glass diffuser. Both 8-bit grayscale and 24-bit RGB OAM-encoded data transfers are carried out in experiments with superior error rates. We can conclude that the OAM-dependent speckles can be not only utilized as a competitive candidate for the OAM mode sorting function in a simple way but also provide an efficient method for the demultiplexing of OAM-encoded data transfer in a practical application.

## 1. Introduction

Optical vortex beams carrying orbital angular momentum (OAM) have been extensively studied in light-matter interaction based applications [1-5]. As a unique optical degree of freedom, the OAMs of the vortex beam are not inherently constrained and thus can be used to increase the channel capacity in optical communication [6-10]. It is worth noting that, in these applications, the measurement and characterization of the complex optical field's OAM spectrum are extremely important, especially in the demultiplexing process in OAM-based data communication systems. Typically, the phase-flattening technique has been recognized as a standard approach to measure and characterize the OAM spectrum, where the incident spatial modes to be evaluated are converted into the fundamental mode through various phase-type components like a spiral phase plate, a fork hologram, or a q-plate [11]. Another kind of typical approach is based on the interferometry technique, where the OAM modes can be effectively separated using cascaded interferometers even at a single-photon level [12-16], or the OAM spectrum can be reconstructed through the inverse Fourier transform of the measured angular coherence function of the complex field using the interferometric method [17, 18]. A spatial mapping technique has also been used to achieve the OAM spectrum. For instance, specific spatial modes can be mapped onto different regions by using a sophisticated hologram [19, 20], or the azimuthal phase front beam can be converted into a beam with a transverse phase gradient by employing a Cartesian to log-polar transformation [21, 22]. In addition, other approaches such as time mapping [23], intensity flattening [24], using a multiplane light conversion (MPLC) device [25, 26], or employing rotational Doppler frequency shift [27] all aim to measure the OAM spectrum. However, the aforementioned techniques suffer from several weaknesses and disadvantages of different kinds, such as poor efficiency (the OAM mode is identified one at a time), requiring fine optical alignment (because of the incorporated interferometers), complex optical configurations (using numerous free-space components, especially for the phase-flattening technique and cascaded interferometers), and limited capability for a large number of spatial components.

On the other hand, the past few decades have witnessed the exciting progress of information extraction from speckle patterns that are achieved after light transmits through a scattering medium [28-31]. Scattering medium such as ground glass or multimode fiber has been employed as a dispersive element for high-resolution spectrometers [32], universal reconfigurable linear operators in optical computation [33], scattering lenses for both imaging through opacity [34, 35] and light focusing [36], and the multispectral imaging with a monochromatic camera [37]. In recent years, many efforts have been dedicated to investigating the diffused light behaviors of optical vortex beams. The random scattering effect has been exploited for the spatial mode sorting assisted by the wavefront shaping technique [38-41]. Besides, the transmission characteristics of Laguerre-Gaussian (LG) [42] and vector [43] vortex beams through turbid media are also investigated. However, in most circumstances, the random behaviors of the vortex beam transmitting through a scattering medium are too intricate to be manipulated and utilized. Commonly used techniques are based on the linear relationship between the incident vortex beam light and the diffused light field after transmitting through the scattering medium. Therefore, the priority is elucidating the random behavior of the considered diffuser, which mainly falls into two typical approaches. One needs to measure the scattering transmission matrix (TM) in the complex field which correlates the incident light with the diffused light [44, 45], while the other one relies on the deep-learning-enabled method by ignoring any phase information and only analyzing the intensities of random speckle patterns [46-49]. However, both the TM calibration and the training process are prohibitively slow and computationally intensive, making them impractical for highly efficient applications.

Here, we focus on exploiting the characteristics of the pure random intensity pattern generated by vortex beams transmitting through a scattering medium. It is theoretically and experimentally revealed that by applying the cross-correlation operation on the OAM-dependent speckles, a correlation annulus is obtained for the first time to the best of our knowledge. It is found that the radius of the cross-correlation annulus is determined by the absolute value of the topological charge difference, which provides an appealing alternative to sorting out the spatial modes. Moreover, by leveraging the orthogonal nature of the OAM-dependent speckles, the incoherent random speckle basis is used to effectively measure the OAM spectrum of a multiplexed coherent light field. The experimentally achieved error rate for an 8-bit grayscale OAM-encoded data transfer is larger than 38 dB for a $100 \times 100$ image with a single basis, while it is ~33 dB for each color channel of a 24-bit RGB OAM-encoded one. The proposed OAM-dependent speckles for spatial mode sorting and multiplexed data transmission can present superior performance than traditional methods, in terms of the cost, complexity, and time consumption in the measurement.

## 2. Results

*2.1 Theoretical analysis of the OAM-dependent speckles*

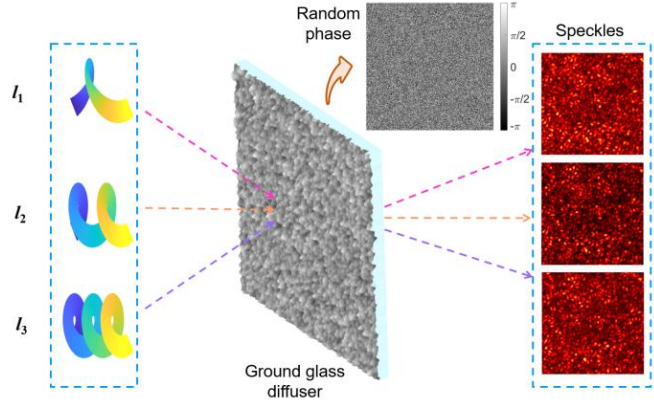

Fig. 1. Schematic diagram of OAM-dependent speckles. Speckles are obtained when vortex beams carrying different topological charges transmit through a fixed ground glass diffuser.

In typical scenarios of the OAM spectrum measurement, the vortex beam sent from the signal transmitter, carrying specific or multiplexed OAM values generally propagates in free space or a waveguide. No appreciable wavefront distortion occurs before the vortex beam arrives at the receiver. Here, we utilize a ground glass diffuser as the effective element for the OAM mode sorting and demultiplexing of the encoded data, as shown schematically in Fig. 1. The thin ground glass diffuser can impose random phase modulation to the incident vortex beam impinged onto it. Then, random speckle patterns can be obtained on the image plane, resulting from the interference between different light paths, which can be described by the theory of Fresnel diffraction, as illustrated in Supplementary Note 1 [50]. Although the highly scrambled light beam is dramatically different from the incident light in terms of the intensity distribution, the effective encoded information is still preserved in the diffused light. This is attributed to the linear light propagation process in a complex scattering medium. The fundamental issue is how to distinguish the OAM component from the speckle patterns without calibrating the TM or applying the deep-learning-based training method.

To concentrate on the measurement of the OAM spectrum in the azimuthal direction and neglect its radial intensity variations for conventional vortex beams (e.g., LG modes) with OAM-dependent radial structure, here we aim to create a perfect vortex beam (PVB) in our experiment based on a Gaussian apodization method as defined in the following model [51, 52]:

$$E_l(r,\varphi) = exp\left(-(r-r_0)^2/\Delta r^2\right) exp(il\varphi) \quad (1)$$

where $r_0$ and $\Delta r$ represent the radius and width of the Gaussian-shaped annulus, respectively, and $l$ denotes the topological charge. Theoretical analysis of the OAM-dependent speckles corresponding to PVBs with topological charges ranging from -20 to 20 are calculated based on the light propagation model for the vortex beam transmitting through scattering medium, as given in Fig. 2. The intensity profiles of the PVBs with different topological charges maintain the same in terms of the radius and width of the annulus, as shown in Fig. 2(a1) for representative PVBs (e.g., $l$=1, 4, 7, and 10). The period of the phase variation corresponds to the value of the topological charge, as shown in Fig. 2(a2). By imposing a fixed random phase modulation on the defined PVBs, scrambled speckle patterns (i.e., $I_l$) are then obtained on the image plane, as given in Fig. 2(a3), which leads to the direct loss of original annulus structural features of PVBs.

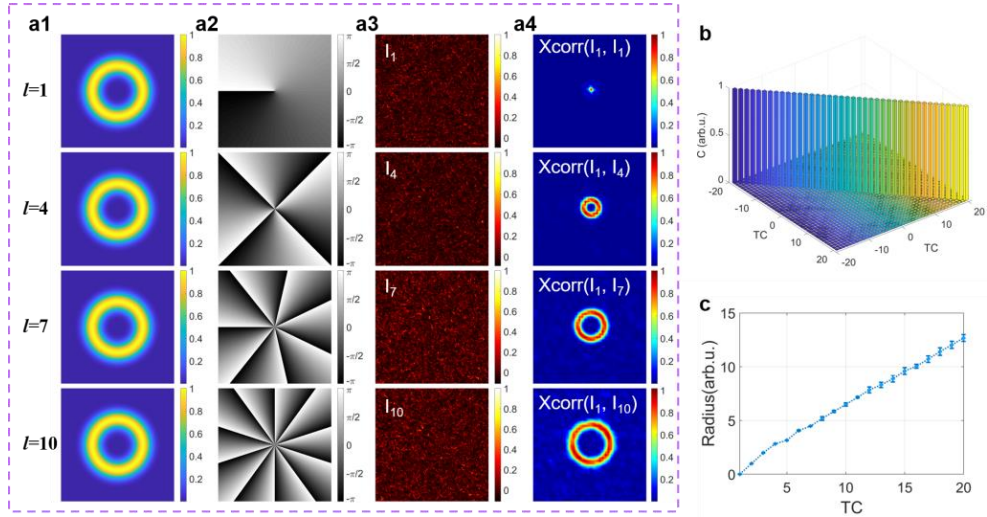

Fig. 2. Theoretical analysis of the OAM-dependent speckles. The intensity (a1) and phase (a2) profiles of the PVBs, and the corresponding OAM-dependent speckles (a3) for a fixed random phase modulation. (a4) Cross-correlation maps between the reference OAM-dependent speckle ($I_1$) and the other ones ($I_1$, $I_4$, $I_7$, and $I_{10}$). Xcorr, cross-correlation. (b) The orthogonal relations between the OAM-dependent speckles corresponding to PVBs with topological charges from $l$=-20 to $l$=20 stepped by 1. (c) The calculated radius of the cross-correlation annulus as a function of topological charge when the reference OAM dependent speckle corresponds to PVB with topological charge $l$=1. TC, topological charge.

First of all, the Pearson correlation coefficients among all OAM-dependent speckles are calculated in Fig. 2(b). The diagonal correlation feature confirms the orthogonality of the OAM-dependent speckles which is inherited essentially from the orthogonality of original PVBs due to the linear transmission through the scattering medium [53]. On the other hand, the orthogonal OAM-dependent speckles are uncorrelated with each other and thus can be used as a set of discrete random bases. In this sense, we can simply assume that the OAM-dependent random bases are mapped from the original OAM ones by utilizing the ground glass diffuser.

Inspired by the works in speckle-correlated imaging [35, 54-56] and optical speckle image velocimetry [57], we resort to the cross-correlation operation on the OAM-dependent speckles to further analyze the effective information concealed in the scrambled speckles. The cross-correlation operation follows the relationship:

$$Xcorr\left(I_{l_1}, I_{l_2}\right) = FT^{-1}\left\{FT\left(I_{l_1}\right)FT^*\left(I_{l_2}\right)\right\} \qquad (2)$$

where $FT$ and $FT^{-1}$ denote the Fourier transform and inverse Fourier transform, $I_l$ is the intensity of the OAM-dependent speckle corresponding to PVB with a topological charge of $l$, and the superscript * represents the conjugate operation. Figure 2(a4) displays the cross-correlation maps between the reference OAM-dependent speckle ($I_1$) and the other ones ($I_1$, $I_4$, $I_7$, and $I_{10}$). The result of cross-correlation between the OAM-dependent speckle and itself, i.e., evolving into the autocorrelation operation like Xcorr($I_1$, $I_1$), is a sharply peaked function as shown in Fig. 2(a4). However, for the OAM-dependent speckles corresponding to PVBs with different topological charges, a clear cross-correlation annulus appears on the cross-correlation maps. Surprisingly, the radius of the cross-correlation annulus seems to be determined by the absolute value of the topological charge difference. To investigate and uncover the possible relationship between the two parameters, we quantitatively calculate the mean value and the standard deviation of the radius of the cross-correlation annulus as a function of topological charge, where the reference OAM-dependent speckle corresponds to $l$=1, as shown in Fig. 2(c). It is observed that the radius of the cross-correlation annulus almost increases monotonically with the topological charge difference. By setting a prior OAM mode with a given topological charge and measuring the corresponding OAM-dependent speckle pattern as a reference, it is theoretically possible to sort the other OAM modes just from the cross-correlation annulus between the OAM-dependent speckles.

*2.2 Experimental results of the OAM-dependent speckles*

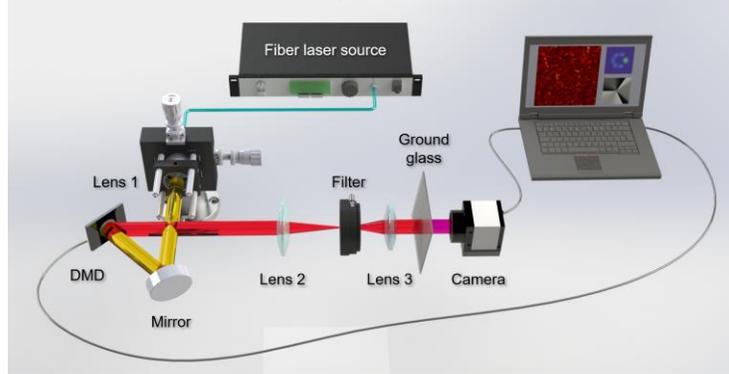

Fig. 3. Experimental diagram of OAM-dependent speckles for spatial mode sorting and multiplexed data transmission. DMD, digital micromirror device.

To experimentally analyze the characteristics of the OAM-dependent speckles, a super-pixel wavefront-shaping technique [58, 59] is employed to generate the PVB and apply different values of OAM onto the beam as schematically shown in Fig. 3. Here, a homemade fiber laser (central wavelength of 1064 nm, 3 dB bandwidth of 0.35 nm) is used as the effective light source for the PVB generation. The output port of the fiber laser source is mounted on a two-dimensional translation stage and further collimated by a convex lens (focal length of 6.2 mm). The collimated light beam is then impinged onto the surface of a digital micromirror device (DMD, DLP650LNIR, Texas Instruments) via a highly reflective mirror. Binary patterns are generated by the super-pixel wavefront-shaping technique and then are loaded onto the DMD. Combined with a 4$f$ imaging system where an iris diaphragm is deployed on the Fourier plane to filter out the 1$^{st}$-order diffraction light, the incident fundamental Gaussian mode is converted into the desired PVBs. The obtained PVBs are then incident onto a piece of ground glass (grit size of 120, Thorlabs), and the OAM-dependent speckles are relayed onto a charge coupled device (CCD, DCU224M, Thorlabs) camera.

First of all, the cross-correlation of the OAM-dependent speckles is experimentally investigated, as shown in Fig. 4(a). Here, the reference OAM-dependent speckle corresponds to PVB with a topological charge of $l=-8$, while the other ones correspond to $l=-8, -5, -2, 2, 5,$ and 8, respectively. Apart from the sharp peak profile shown in the autocorrelation map of $l=-8$, a cross-correlation annulus can be discerned on the cross-correlation map, the radius of which depends on the topological charge difference. This agrees well with the theoretical analysis in Fig. 2(a4). The gradually pronounced background noise in the cross-correlation map can be attributed to the low-pass filtering effect of the ground glass diffuser. To further explain the underlined mechanism that the OAM-dependent speckles can work for the OAM mode sorting, a proof-of-principle experiment is provided in Supplementary Note 2, illustrating the evolution features of the cross-correlation map among OAM-dependent speckles corresponding to PVBs with topological charges ranging from $l=-3$ to $l=3$. It is worth acknowledging that the cross-correlation annulus cannot give the true value of the topological charge. However, since its cross-correlation annulus is only determined by the absolute value of the topological charge difference, it is sufficient to sort the OAM modes especially when the detecting modes are well-defined and the reference OAM-dependent speckle is known as prior information. In contrast to traditional OAM spectrum measurement approaches, the cross-correlation of the OAM-dependent speckles greatly reduces the system complexity in terms of the amount of required optical components (just a piece of ground glass diffuser here) and stability (no need for strict interferometric alignment). The calibration only necessitates a prior OAM-dependent speckle, which is easier to implement and much more stable compared with the complex field TM measurement and the training process in the deep-learning approach.

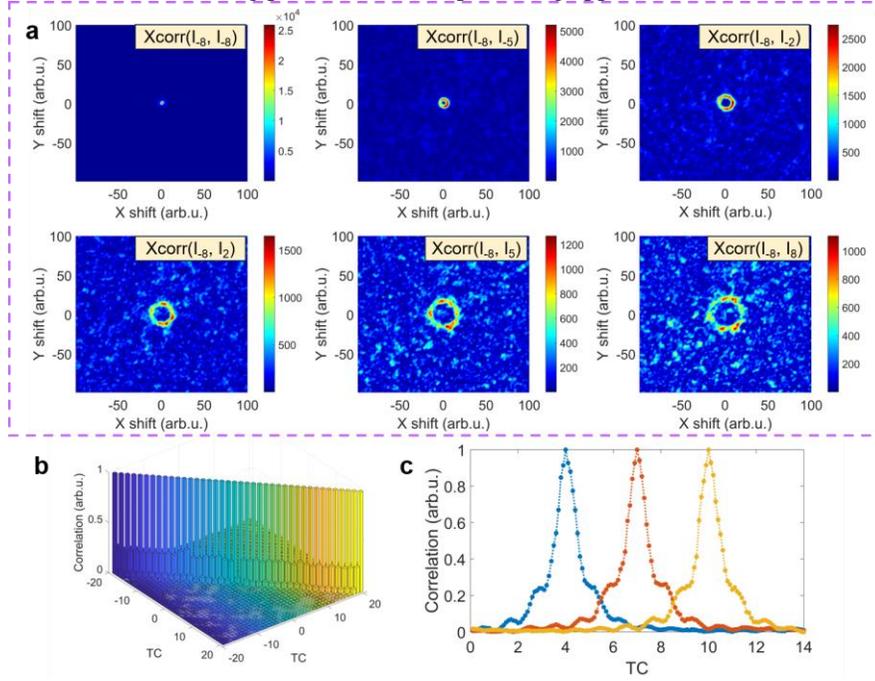

Fig. 4. Experimental analysis of the OAM-dependent speckles. (a). The measured cross-correlation maps between the reference OAM-dependent speckle ($I_{-8}$) and the other ones ($I_{-8}$, $I_{-5}$, $I_{-2}$, $I_2$, $I_5$, $I_8$). (b) The measured orthogonal relations between the OAM-dependent speckles corresponding to PVBs with topological charges ranging from $l=-20$ to $l=20$ stepped by 1. (c) OAM-dependent speckle decorrelation bandwidth.

The orthogonal relation among the OAM-dependent speckles corresponding to PVBs with topological charges ranging from $l=-20$ to $l=20$ is also experimentally verified by the diagonal correlation feature, as indicated in Fig. 4(b). It is also shown that the correlation coefficient between the neighboring integer orders is around 0.25, which can be regarded as decorrelated speckles. To give a more quantitative characterization of the decorrelation bandwidth (defined by the change in topological charge to reduce the correlation coefficient by half), the correlation coefficients of the OAM-dependent speckles corresponding to PVBs with topological charge centered at $l=4$, 7, and 10 stepped by 0.1 are provided in Fig. 4(c). The OAM-dependent speckle decorrelation bandwidth is found to be ~0.5. Therefore, for a discrete decorrelated random speckle basis, an integer order based topological charge separation is enough for the decorrelation of the OAM-dependent speckles and guarantees efficient demultiplexing.

*2.3 Demultiplexing of OAM-encoded data transfer using scattering medium*

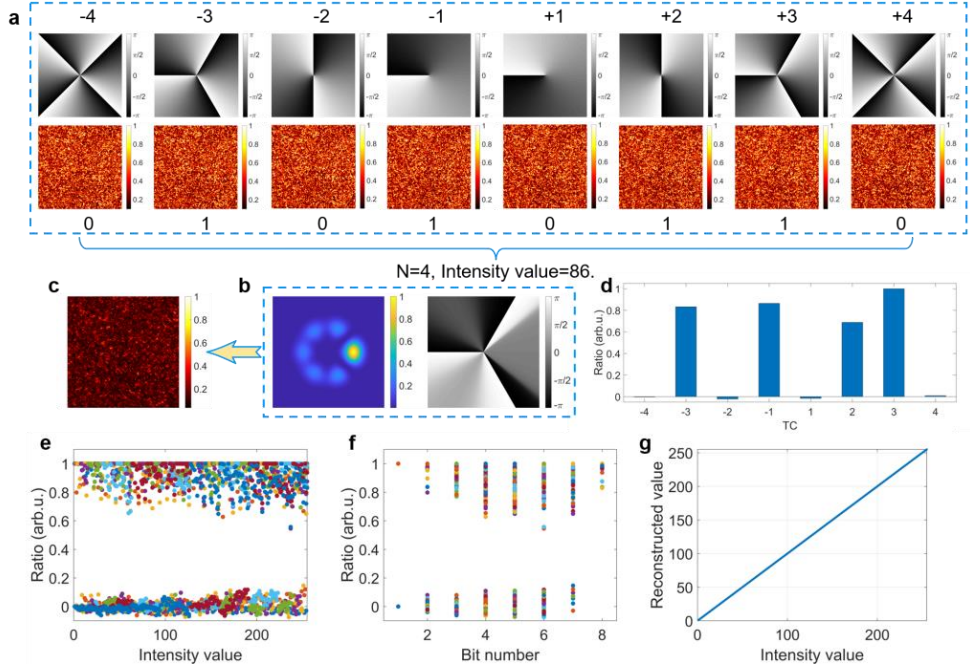

Fig. 5. An 8-bit grayscale OAM-encoded data transfer which is demultiplexed by the OAM-dependent speckle basis. (a) 8-OAM bases and the corresponding OAM-dependent speckles. (b) The multiplexed intensity and phase profiles correspond to a grayscale intensity value of 86. (c) A speckle pattern obtained by transmitting the multiplexed data through a ground glass diffuser. (d) Experimental OAM spectrum of intensity value 86 demultiplexed by the OAM dependent speckle basis. Statistical analysis of OAM spectra as a function of gray scale pixel value (e) and bit number (f). (g) Reconstructed intensity value versus the ground truth intensity value.

For the purpose of optical communication, OAM-encoded data from the super-pixel wavefront shaping is generated and the OAM-dependent speckle basis is employed as the key for demultiplexing. For an 8-bit grayscale OAM-encoded data transfer, the OAM basis composed of 8 PVBs (i.e., $l=\pm 4, \pm 3, \pm 2,$ and $\pm 1$) is employed for the data encoding, as defined in Fig. 5(a). For each PVB basis, the OAM-dependent speckle, i.e., the OAM-dependent transmission function $H(x, y, l)$, is correspondingly measured. The 8-bit grayscale intensity value, i.e., $O(l)$, is then encoded, representing 256 gray levels. For example, the grayscale intensity value 86 is transformed into the binary byte of '01010110' with 4 1-value bits (i.e., N=4). Figure 5(b) gives the theoretical intensity and phase profiles of the encoded beam. Taking advantage of the decorrelation characteristics of the ground glass diffuser, the measured speckle pattern $I(x, y)$ obtained after the encoded signal transmitting through the ground glass diffuser can be written as:

$$I(x, y) = \sum_{l} H(x, y, l) O(l) \tag{3}$$

Therefore, the encoded signal $O(l)$ can be demultiplexed through a pseudoinverse operation, and the reconstructed OAM spectrum is shown in Fig. 5(d). Here, the reconstructed power ratio of each OAM channel is normalized by the maximum value of the original calculated ones to promote the available threshold range for binarization. From the calculated OAM spectrum in Fig. 5(d), it is clear that the correct binarization can be realized by setting the threshold between 0.009 to 0.687. To further investigate the statistical characteristics of the OAM-dependent speckle basis for data demultiplexing, grayscale intensity values from 1 to 255 are considered. The normalized power ratio of the calculated OAM spectra as a function of intensity values and bit numbers are shown in Fig. 5(e) and 5(f), respectively. The threshold range between 0.147 to 0.545 is clearly shown in both results, which is still a very broad range for efficient binarization. By setting any threshold values within this range, the reconstructed intensity values versus the ground truth intensity values are obtained in Fig. 5(g). The smooth and linear profile verifies the 100% correct intensity value reconstruction demultiplexed by the OAM-dependent speckle basis.

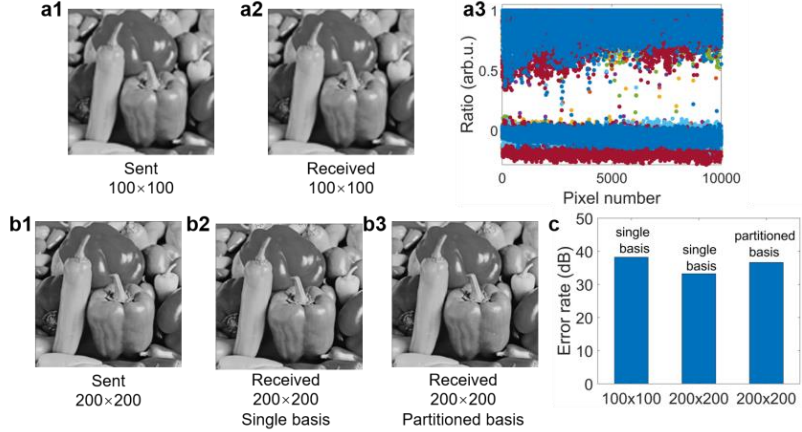

Fig. 6. Results of the 8-bit grayscale OAM-encoded data transfer. Sent (a1) and received (a2) grayscale images (Peppers, 100 × 100 pixels). (a3) Statistical analysis of OAM spectra as a function of pixel number. Sent (b1) and received grayscale images (Peppers, 200 × 200 pixels) with single (b2) or partitioned (b3) basis. (c) Error rates of the results in (a2), (b2) and (b3).

Leveraging the above OAM encoding and decoding strategy, an 8-bit grayscale image (Peppers [60], $m \times n$ pixels) is used for data transfer, as shown in Fig. 6. The image is sent and decoded pixel by pixel which can be written as:

$$I_{xy \times mn} = \sum_{l} H_{xy \times l} O_{l \times mn} \tag{4}$$

where $I_{xy \times mn}$ is reshaped from $mn$ speckle patterns (each of which is $x \times y$ in size) corresponding to the image size of $m \times n$, $H_{xy \times l}$ is reshaped from $l$ OAM-dependent speckle basis, i.e., $l=8$ for 8-bit grayscale data transfer, and $O_{l \times mn}$ is the target OAM spectrum to be reconstructed. Here, a $100 \times 100$ pixels sized image (Fig. 6(a1)) is first considered. Statistical analysis of the OAM spectra as a function of the pixel number is displayed in Fig. 6(a3). It is worth noting that the OAM-dependent speckle basis $H_{xy \times l}$ is measured in the middle when sending the $100 \times 100$ speckle images to suppress the decorrelation effect of the OAM-dependent speckles. It is found that the clear band gap of the threshold selection range may become narrowed for the speckle patterns measured with a larger time interval with respect to the OAM-dependent speckle basis. The scatters within the band gap region would lead to inaccurate demultiplexing. Notwithstanding, high-quality image reconstruction is realized in Fig. 6(a2) after taking the binarization operation. The error rate of this 8-bit grayscale OAM-encoded data transfer is calculated to be 38.24 dB with only one set of OAM-dependent speckle bases, as shown in Fig. 6(c).

To verify the capability of data transfer in a larger size, the image of interest is increased to 200×200 pixels as shown in Fig. 6(b1). Here, two decoding strategies are employed respectively, one is using a single set of OAM-dependent speckle bases for the whole speckle images, and the other one is dividing the 200×200 image into 5 groups, using partitioned basis to reconstruct each part and combing the reconstructed parts together as a whole, as shown in Figs. 6(b2) and 6(b3). We can see an obvious decline in the error rate for the first single basis strategy, i.e., 33.23 dB, and the error rate value returns to 36.66 dB for the partitioned basis, as shown in Fig. 6(c). Both reconstructed images show high fidelity in contrast with the ground truth image. To further investigate the stability of the system that is closely related to the demultiplexing efficiency, the speckle decorrelation effect is evaluated in detail in Supplementary Note 3.

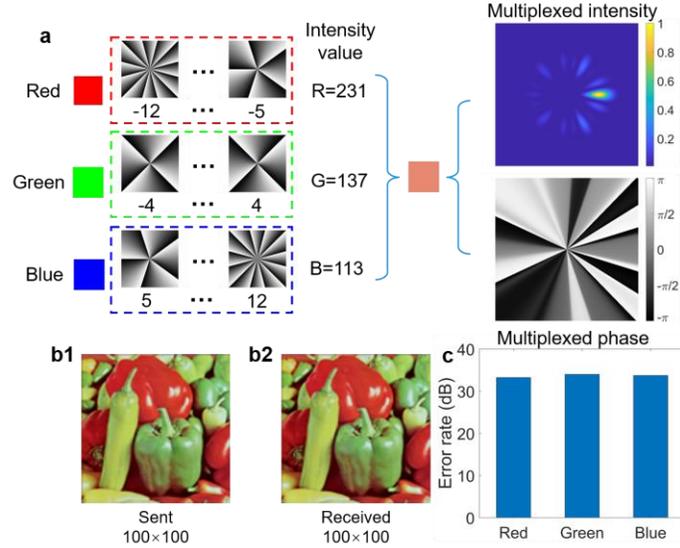

Fig. 7. Results of the 24-bit RGB OAM-encoded data transfer. (a) Schematic diagram of the 24-bit RGB encoding scheme using 24-OAM multiplexing. Sent (b1) and received (b2) RGB color images (Peppers, 100 × 100 pixels). (c) Error rates in each RGB channel.

Finally, the concept of OAM-dependent speckles in the data transfer application is further extended to a color image that is weighted by three primary colors, i.e., red, green, and blue. The encoding scheme of the 24-bit RGB data is provided in Fig. 7(a). The PVBs with topological charges ranging from -12 to 12 are considered as the basis, where the red color takes the topological charges from -12 to -5, the green one from -4 to 4, and the blue one from 5 to 12, respectively. Therefore, each primary color layer is composed of an 8-bit grayscale image and the mixture of the three layers yields the RGB color image. Figure 7(a) shows the multiplexed intensity and phase profiles of a specific color, i.e., red, 231; green, 137; blue, 113. It is worth noting that the considered image is 100×100 and only one set of OAM-dependent speckle bases is employed. The reconstructed RGB color image in Fig. 7(b2) matches well with the ground truth image in Fig. 7(b1), except for a small number of visible noise points. The error rates in each of the RGB channels, i.e., 33.23 dB for red, 33.98 dB for green, and 33.72 dB for blue as shown in Fig. 7(c), also reflect the high demultiplexing efficiency using the OAM-dependent speckle basis even it is extended to a 24-bit RGB OAM-encoded data.

*2.4 Discussions*

Compared with traditional methods for the OAM mode sorting, the cross-correlation approach takes full advantage of the scattering effect to achieve OAM-dependent speckles, which requires only a piece of ground glass as the key element. Besides, the cross-correlation calculations are based on the speckle patterns which are incoherently measured after the vortex beam transmits through the ground glass diffuser, without using any sophisticated optical components (such as spiral phase plate, fork hologram, or q-plate), or complex optical alignment (such as in cascaded interferometers). With prior information on a specific OAM-dependent speckle pattern for calibration, the OAM mode sorting can be realized by characterizing the radius of the cross-correlation annulus. No further characterization of the scattering diffuser is needed, such as tailoring the scattering effect actively by the wavefront shaping or the coherently calibrated TM using the interferometric technique. In addition, the cross-correlation operation of the speckle patterns has also been widely used in speckle-correlated imaging and speckle image velocimetry and can be efficiently calculated.

More importantly, the OAM-dependent speckles corresponding to the vortex beam with different topological charges are used as a set of OAM-dependent speckle bases for the demultiplexing of OAM-encoded data. Distinct from commonly used approaches where the TM of the scattering diffuser is calibrated, or the speckles are thoroughly analyzed with the assistance of the deep-learning method, the demultiplexing technique proposed here only requires the calibration of the OAM-dependent speckle basis (i.e., 8(24) speckle patterns for 8(24)-bit OAM-encoded data). The latter is much simpler than the calibration of a traditional TM or the training process in a deep-learning approach. For a 100×100 sized image, the required measurement of the OAM-dependent speckle basis for the 8(24)-bit OAM-encoded data only accounts for 0.08% (0.24%), which takes up a relatively small portion of the whole data volume. A ground glass diffuser is the only key element that is required, greatly reducing the demultiplexing complexity for the OAM-encoded data.

It is also worth noting that the rate of OAM-encoded data transfer using the OAM-dependent speckle basis is mainly restricted by the frame rate (i.e., 15 fps here) of the CCD camera. It is speculated that by using a high-speed camera, the OAM-encoded data acquisition rate can be greatly accelerated. In addition, the speckle decorrelation effect can be further suppressed to reduce the error rate by fixing the ground glass diffuser and the camera together to reduce the relative shift from environmental vibrations, especially when a much larger data volume is sent and demultiplexed.

## 3. Conclusion

In conclusion, the characteristics of OAM-dependent speckles, originating from the vortex beam transmitting through a ground glass diffuser, are theoretically and experimentally investigated. It is found that, by imposing a cross-correlation between the OAM-dependent speckles, an annulus can be obtained on the cross-correlation map with its radius dependent on their topological charge difference. Therefore, the OAM modes can be effectively sorted with prior information on a specific OAM-dependent speckle as a reference. Furthermore, an OAM-dependent speckle basis can be also utilized as a competitive candidate for the demultiplexing of OAM-encoded data transfer. The comparatively low error rates of both 8-bit grayscale and 24-bit RGB OAM-encoded data transfers indicate the feasibility of the OAM-dependent speckle-based demultiplexing. This work reveals the fundamental features of OAM-dependent speckles and paves an appealing method for OAM-encoded data transfer in practical applications.

**Funding.** Guangdong Basic and Applied Basic Research Foundation (Grant No. 2020A1515111143); Natural Science Foundation of Guangdong Province (Grant No. 2021A1515011532); Shenzhen Government's Plan of Science and Technology (Grant No. JCYJ20220818100019040, RCYX20210609103157071).

**Acknowledgments.** R. M. conceived the idea and wrote the original manuscript. R. M. and K. H. L. conducted the experiments. K. H. L. realized the software control and data measurement. R. M. and J. L. funded the work. J. L. and J. S. H. supervised the project. Z. W., W. L. Z., D. Y. F., and A. S. L. G. discussed the results. All the authors discussed the results and contributed to the writing of the manuscript.

**Disclosures.** The authors declare no conflicts of interest.

**Data availability.** Data underlying the results presented in this paper are not publicly available at this time but may be obtained from the authors upon reasonable request.

**Supplemental document.** See Supplementary for supporting content.